\documentclass[aps,prd,twocolumn,showpacs,amsmath,nofootinbib]{revtex4-1}

\usepackage{color}
\newcommand{\beqa}{\begin{eqnarray}}
\newcommand{\eeqa}{\end{eqnarray}}
\usepackage{float}
\usepackage{epsfig}
\usepackage{graphicx}
\usepackage{latexsym}
\usepackage{amsmath}
\usepackage{hyperref}
\usepackage{mathrsfs}
\usepackage{dcolumn}
\usepackage{bm}%

\begin{document}

\title{Muon neutrinos and the cosmological abundance of primordial black holes}

\author{Jiali Hao$^1$}
\author{Yupeng Yang$^{1,2}$}\email{ypyang@aliyun.com}
\author{Qianyong Li$^1$}
\author{Yankun Qu$^1$}
\author{Shuangxi Yi$^1$}
\affiliation{$^1$School of Physics and Physical Engineering, Qufu Normal University, Qufu, Shandong 273165, China \\
             $^2$Joint Center for Particle, Nuclear Physics and Cosmology, Nanjing University-Purple Mountain Observatory, Nanjing, Jiangsu 210093, China}

\begin{abstract}

In the mixed dark matter scenarios consisting of primordial black holes (PBHs) and particle dark matter (DM), PBHs can accrete surrounding DM particles to form ultracompact minihalos (UCMHs or clothed PBHs) even at an early epoch of the Universe. The distribution of DM particles in a UCMH follows a steeper density profile ($\rho_{\rm DM}\propto r^{-9⁄4}$) compared with a classical DM halo. It is expected that the DM annihilation rate is very large in UCMHs, resulting in a contribution to, e.g., the extragalactic neutrino flux. In this work, we investigate the extragalactic neutrino flux from clothed PBHs due to DM annihilation, and then the muon flux for neutrino detection. Compared with the atmospheric neutrino flux, we derive the upper limits on the cosmological abundance of PBHs for 10 years of exposure time of, e.g., the IceCube experiment. For the DM mass $m_{\chi}=100~(1000)$ GeV, the upper limits (2$\sigma$) on the fraction of DM in PBHs are $f_{\rm PBH}=1.2\times 10^{-3}~(8.9\times 10^{-5})$ for contained events and $f_{\rm PBH}=2.5\times 10^{-3}~(1.3\times 10^{-5})$ for upward events. Compared with other constraints, although the upper limits obtained by us are not the strongest, it is a different way to study the cosmological abundance of PBHs.

\end{abstract}

\maketitle

\section{introduction} 

Dark matter (DM) is a fundamental challenge in modern astrophysics and theoretical physics, and the study of it heralds a profound revolution in our comprehension of the Universe~\cite{2005PhR...405..279B,Jungman:1995df}. In the standard cosmological model, DM accounts for $\sim 26\%$ of the total energy density of the Universe~\cite{2020A&A...641A...6P}. Although the existence of DM has been confirmed by many astronomical observations, our understanding of its nature remains limited. Existing astronomical observations and related studies show that DM is not involved in, e.g., electromagnetic interactions; therefore, it is difficult to detect DM directly. However, the existence of DM can be inferred from its gravitational effect on visible objects~\cite{2005PhR...405..279B, 2020A&A...641A...6P, 2021arXiv210905854G}.

Many DM models have been proposed, of which the weakly interacting massive particles (WIMPs) model based on supersymmetric string theory is the most widely studied~\cite{2005PhR...405..279B,Jungman:1995df}. According to the related theory, WIMPs can undergo annihilation processes resulting in the production of standard model particles such as photons, electrons/positrons, and neutrinos ~\cite{2005PhR...405..279B, 1996PhR...267..195J,2017MPLA...3230022A}. Consequently, detecting these produced particles through astronomical observations provides an indirect way of probing the nature of particle DM. Since the annihilation rate of DM is proportional to the square of the number density, it is a good choice to conduct relevant studies in regions with large particle density~\cite{2011PhRvD..84d3506Y, 2011EPJP..126..123Y, 2011JCAP...12..020Y,Hooper:2010mq,PhysRevLett.93.241301,2022JCAP...08..065F}.

It is well known that the current large-scale structure of the Universe results from the evolution of early density perturbation with an amplitude of $\delta \rho/\rho\sim10^{-5}$~\cite{2011ApJS..192...18K}. If there are large density perturbations in the early universe, e.g., $\delta \rho/\rho\gtrsim 0.3$, they would collapse directly to form primordial black holes (PBHs)~\cite{1997PhRvD..56.6166G,2021RPPh...84k6902C,2023PDU....4101231B,Heydari:2021gea,Heydari:2021qsr,Heydari:2023xts,Heydari:2023rmq}. PBHs spread over a wide mass range, and different astronomical observations have been used to constrain their cosmological abundance~\cite{1997PhRvD..56.6166G,2021RPPh...84k6902C, 2023PDU....4101231B,2022arXiv220514722O, 2022PhRvD.106d3516Y,2020EPJP..135..690Y, 2022ApJ...928L..13Z, 2023EPJC...83..934Y, Su:2024hrp,2021RPPh...84k6902C, Auffinger:2022khh, Bernal:2022swt, Berteaud:2022tws, Coogan:2020tuf, Tan:2022lbm, Facchinetti:2022kbg, Xie:2024eug, 2010ApJ...720L..67L, 2022JHEP...12..090C, 2024arXiv240411053Z, 2022MNRAS.511.1141Z, 2022MNRAS.513.3627Z, 2022PhRvD.106d3003K, 2022PhRvD.106e5043L, 2022PhRvD.105f3008C, 2024MNRAS.tmp.1099T, 2022MNRAS.517.1086Z, 2022PhRvD.105h3526F, 2021PhRvD.103d3010W, 2022PhRvD.106i5034L, 2021JCAP...05..051C, 2021JPhG...48d3001G,Andres-Carcasona:2024wqk,Zhang:2023rnp,Cang:2020aoo,Clark:2018ghm,Cai:2022kbp}. 
On the other hand, if the amplitude of early density perturbation is in the range of $10^{-5}\lesssim \delta \rho/\rho\lesssim 0.3$, it has been proposed that a kind of DM structure, ultracompact minihalos (UCMHs), can be formed in the early universe~\cite{2009ApJ...707..979R}. The density profile of DM particles in UCMHs is $\rho_{\rm DM}\propto r^{-9⁄4}$~\cite{2009ApJ...707..979R,2020PhRvR...2b3204S}, which is significantly steeper than that of the classical DM halo model, e.g., the Navarro-Frenk-White model. 
Therefore, the annihilation rate of DM particles in UCMHs will be large, and they can be referred to as a kind of potential high-energy astrophysical source~\cite{2013PhRvD..87j3525Y,2017PhRvD..96j3509Y,2022PhRvD.105d3011Z,Scott:2009tu,PhysRevD.85.125027,Gines:2022qzy}. 

However, the numerical simulations have shown that, after the formation of UCMHs, the final density profile of DM particles is not so steep due to the effects of evolution~\cite{2017PhRvD..96l3519G,2018PhRvD..97d1303D}. Therefore, it would be difficult to form UCMHs with an unchanged density profile of $\rho_{\rm DM}\propto r^{-9⁄4}$ via the direct collapse of early large density perturbations. On the other hand, it has been proposed that UCMHs with an unchanged density profile can be formed through the accretion of DM particles onto  PBHs~\cite{2021PhRvD.103l3532T,2019PhRvD.100b3506A,Boucenna:2017ghj}. Therefore, in mixed DM scenarios, if DM is partly comprised of WIMPs and partly of PBHs, the detectability of WIMPs is enhanced due to its accretion onto PBHs. In previous works, e.g., Refs.~\cite{2022JCAP...08..065F,2010ApJ...720L..67L}, this has been investigated in the context of electromagnetic signatures of WIMP annihilation. In this work, we will mainly focus on neutrino signatures at IceCube.

Since DM particles in UCMHs can annihilate into high-energy photons or electrons/positrons, it is possible to investigate the cosmological abundance of PBHs by studying relevant astronomical observations~\cite{2020EPJP..135..690Y, 2011JCAP...12..020Y, 2022PhRvD.105d3011Z, 2021PhRvD.103l3532T, 2021MNRAS.506.3648C,Cai:2020fnq}. In addition to photons, DM particles can also annihilate into neutrinos, and they are almost unaffected by the medium during their propagation compared to photons and charged particles. Moreover, the study of neutrinos provides a good complement to researching the nature of particle DM with large mass~\cite{2021RvMP...93c5007A}. In Ref.~\cite{2010ApJ...720L..67L}, the authors investigated the potential neutrino signals from DM spikes surrounding black holes in our Galaxy. The authors of~\cite{2013PhRvD..87j3525Y,2017PhRvD..96j3509Y} studied the neutrino signals from nearby UCMHs due to DM annihilation, and obtained the constraints on the abundance of UCMHs.~\footnote{In Refs.~\cite{2013PhRvD..87j3525Y,2017PhRvD..96j3509Y}, the authors adopted that UCMHs 
are formed via the collapse of large density perturbation in the early universe. Therefore, they finally obtained the cosmological abundance of UCMHs.} 
Since UCMHs can be formed in the early universe, it is expected that there should be extragalactic neutrinos produced by the annihilation of DM particles in UCMHs, and we focus on this issue in this work. We adopt that UCMHs are formed by accreting surrounding particle DM onto PBHs.~\footnote{Note that in this scenario PBHs should not make up all the components of DM ($\Omega_{\rm PBH}< \Omega_{\rm DM}$).}  

This paper is organized as follows. In Sec.~\ref{THE DENSITY OF PBHS}, we briefly review the basic properties of UCMHs. In Sec.~\ref{neutrinos flux}, we investigate the neutrino flux from PBHs and the muon flux for neutrino detection. In Sec.~\ref{The constraints on the abundance of PBHs}, we derive the upper limits on the cosmological abundance of PBHs using the neutrino signals, and then the conclusions are given in Sec.~\ref{conclusions}. 

\section{the basic properties of UCMHs}
\label{THE DENSITY OF PBHS}

In the mixed DM scenarios consisting of WIMPs and PBHs, WIMPs could be accreted onto PBHs to form UCMHs. The mass of a UCMH increases very slowly until the redshift at $z_{\rm eq}\sim 3400$~\cite{Scott:2009tu}. Theoretical research and numerical simulations have shown that the density profile of DM particles in a UCMH is 
$\rho_{\rm DM }\propto r^{-9/4}$. On the other hand, considering the annihilation of DM particles, there is a maximum value 
$\rho_{\rm core}$ at the center of UCMHs. Therefore, the density profile of a UCMH at redshift $z$ can be written as~\cite{2021PhRvD.103l3532T}

\beqa
\rho(r,z)=\left\{
\begin{array}{rcl}
\rho_{\rm core} ,\,\,\,\,\,\,\,\,\,\,\,\,\,\,\,\,\,\,\,\,\,\,\,\,\,\,\,\,\,\,\,\,\,\,\,\,\,\,\,\, &&{r< r_{\rm cut}(z)}\\
\\
\rho_{\rm core}(r/r_{\rm cut}(z))^{-9/4}, && { r_{\rm cut}(z)\leq r<r_{\rm ta}(z_{\rm eq})}\\
\end{array} \right.
\eeqa
where $r_{\rm ta}(z)$ is the turnaround scale at redshift $z$~\cite{2019PhRvD.100b3506A}, 
\beqa
r_{\rm ta}(z) \approx (2GM_{\rm PBH}t(z)^2)^{1/3}.
\label{eq:r_ta}
\eeqa

The center density $\rho_{\rm core}$ depends on the properties of particle DM, and it can be written as~\cite{Ullio:2002pj,2021PhRvD.103l3532T},

\beqa
\rho_{\rm max} = \frac{m_{\chi}}{\left<\sigma v\right>(t(z)-t_{i})}
\label{eq:rho_max}
\eeqa
where $t_{i}$ is the formation time of UCMHs, $m_{\chi}$ is the mass of DM particles,  $\left<\sigma v\right>$ is the thermally averaged annihilation cross section, and we adopt a benchmark value of $\left<\sigma v\right> =3\times10^{-26}\, \rm cm^3s^{-1}$~\cite{2012PhRvD..86b3506S}. $r_{\rm cut}$ can be determined by the condition of $\rho_{\rm core}=\rho(r_{\rm cut})$, 

\beqa
r_{\rm cut}(z) = \left( \frac{\rho_{\rm max}}{\bar{\rho}_{\rm DM}(z_{\rm eq})}\right)^{-9/4}r_{\rm ta}(z_{\rm eq}).
\label{eq:rho_max}
\eeqa
where $\bar{\rho}_{\rm DM}(z_{\rm eq})$ is the mean density of DM at $z_{\rm eq}$. 
Ignoring the kinetic energy of particle DM compared to the potential energy, we consider the mass of PBHs in the following range~\cite{2021PhRvD.103l3532T}, 

\beqa
\begin{split}
M_{\rm PBH}  &\ge6.5\times 10^{-4} M_{\odot } \\&\times \left (\frac{\left <\sigma v  \right > }{3\times10^{-26}\, \rm cm^3s^{-1}}\right )^{-1/3} \left ( \frac{m_{\chi } }{10 \,\rm GeV}  \right ) ^{-73/24}.   
\label{eq:M_{PBH}}
\end{split}
\eeqa

\section{extragalatic neutrino flux from dark matter annihilation in UCMHs and muon flux for detection}
\label{neutrinos flux}

Previous works have mostly focused on the extragalactic or galactic gamma-ray flux from DM annihilation in UCMHs~\cite{2011JCAP...12..020Y,2020EPJP..135..690Y,2022PhRvD.105d3011Z,PhysRevD.85.125027,Scott:2009tu,2021MNRAS.506.3648C}. The authors of~\cite{2013PhRvD..87j3525Y} investigated the neutrino flux from nearby UCMHs. Here we will focus on the extralgalactic neutrino flux from DM annihilation in UCMHs. Similar to the extragalactic gamma-ray flux, the differential neutrino flux can be written as,

\begin{equation}
\begin{split}
\frac{d\phi_{\rm \nu}}{dE_{\rm \nu}}=&\frac{\Omega_{\rm PBH}\rho_{\rm c,0}}{M_{\rm PBH}}\frac{c}{8\pi}\frac{\left<\sigma v\right>}{m_{\rm \chi}^2}\\&\times\int_{0}^{z_{\rm up}}\,\frac{dz}{H(z)}\frac{dN_{\rm \nu}}{dE_{\rm \nu}}(E^\prime,z)\int\rho^2(r,z)dV,
\label{eq:neutrinos_flux}
\end{split}
\end{equation}
where $\Omega_{\rm PBH}=\rho_{\rm PBH,0}/\rho_{c,0}$ is the abundance of PBHs, $E^\prime = E(1 + z)$ and $z_{\rm up} = m_{\chi}/E-1$. $dN_{\rm \nu}/{dE_{\rm \nu}}$ is the energy spectrum of neutrinos and it can be obtained using the public code, e.g., $\mathtt{DarkSUSY}$~\cite{Bringmann:2018lay,Gondolo:2004sc}.~\footnote{https://darksusy.hepforge.org/}

In the standard model, there are three flavors of neutrinos, $\nu_e(\bar{\nu}_e)$, $\nu_\tau(\bar{\nu}_\tau)$ and $\nu_\mu(\bar{\nu}_\mu)$. Due to the neutrino oscillation in a vacuum, they can convert into each other during their propagation. In this work, for simplicity, we set the ratio between the neutrino flavors as 1:1:1. Neutrinos generated by DM annihilation in UCMHs do not lose their energy during their propagation to the Earth. When they reach the side of Earth, muon neutrinos ($\nu_{\mu}$) can be converted into muons ($\mu$) through charged current interaction with matter (e.g., rock or ice). These muons can be detected by the detector on Earth via, e.g., Cherenkov light. In view of the muon detection, for our purposes, here we consider two popular kinds of events named "upward events" and "contained events". 

For contained events, the muons are produced in the detector through the charged current interaction, and the differential muon flux can be written as~\cite{2009PhRvD..80d3514E},

\beqa
\begin{split}
\frac{d\phi_{\rm \mu}}{dE_{\rm \mu}}\bigg|_{\rm con}=&\frac{N_{A}\rho}{2}\int_{E_{\rm \mu}}^{m_{\rm \chi}}\,dE_{\rm \nu}\left(\frac{d\phi_{\rm \nu}}{dE_{\rm \nu}}\right)\\ &\times\left(\frac{d\sigma_{\rm \nu}^p(E_{\rm \nu},E_{\rm \mu})}{dE_{\rm \mu}}+(p\to n)\right)+(\nu \to \bar{\nu}),
\label{eq:flux_ct}
\end{split}
\eeqa 
where $\rho$ is the density of the medium, $N_{A}=6.022\times10^{23}$ is Avogadro's number. $\frac{d\sigma_{\rm \nu,\bar{\nu}}^{\rm p,n}}{dE_{\rm \mu}}$ are the scattering cross sections of neutrinos (antineutrinos) off protons and neutrons, and here we adopt following form~\cite{2006hep.ph....6054S, 2007PhRvD..76i5008B, 2009PhRvD..80d3514E}

\beqa
\begin{split}
\frac{d\sigma_{\rm \nu,\bar{\nu}}^{\rm p,n}}{dE_{\rm \mu}}=\frac{2{m_p}{G_F^2}}{\pi}\left(a_{\rm \nu,\bar{\nu}}^{\rm p,n}+b_{\rm \nu,\bar{\nu}}^{\rm p,n}\frac{E_{\mu}^2}{E_{\nu}^2}\right),
\label{eq:m_ucmh}
\end{split}
\eeqa 
where $a_{\rm \nu}^{\rm p,n}=0.15, 0.25$, $b_{\rm \nu}^{\rm p,n}=0.04, 0.06$ and $a_{\bar{\nu}}^{\rm p,n}=0.06, 0.04, b_{\bar{\nu}}^{\rm p,n}=0.25, 0.15.$ What the above formula describes are the weakly  scattering charged-current cross sections of $\nu_\mu(\bar{\nu}_\mu)$ scattered with protons and neutrons.

For upward events, the muons are produced outside the detector when neutrinos arrive on the opposite side and travel through the interior of Earth. The differential muon flux can be written as~\cite{2009PhRvD..80d3514E},

\beqa
\begin{split}
\frac{d\phi_{\rm \mu}}{dE_{\rm \mu}}\bigg|_{\rm up}=&\frac{N_{A}\rho}{2}\int_{E_{\rm \mu}}^{m}\,dE_{\rm \nu} \left(\frac{d\phi_{\rm \nu}}{dE_{\rm \nu}}\right)\\ & \times\left(\frac{d\sigma_{\rm \nu}^p(E_{\rm \nu},E_{\rm \mu})}{dE_{\rm \mu}}+(p\to n)\right)R(E_{\rm \mu})+(\nu \to \bar{\nu}),
\label{eq:flux_up}
\end{split}
\eeqa
where $R(E_{\rm \mu})$ is the distance of the muons traveled in the medium until their energy falls below the threshold of the detector~\cite{2010PhRvD..82b3506Y, 2010PhRvD..81h3506S}, and can be written as,

\beqa
R(E_{\rm \mu})=\frac{1}{\rho \beta}ln\frac{\alpha+\beta E_{\rm \mu}}{\alpha+\beta E_{\rm \mu}^{\rm th}},
\label{eq:rmu}
\eeqa
where $\alpha \sim10^{-3}\, {\rm GeV{cm}^2 g^{-1}}$ relates to the ionization energy loss, and $\beta \sim10^{-6}\,{\rm cm}^2{\rm g}$ denotes the loss of other radiation energy due to bremsstrahlung and pair creation. $E_{\rm \mu}^{\rm th}$ represents the threshold energy of the detector, and here we set $ E_{\mu}^{th} = 50$ GeV.

For detecting neutrinos, the main background is the atmospheric neutrinos (ATM), and it has been detected by related experiments~\cite{2002ARNPS..52..153G, 2007PhRvD..75d3006H}. For our purposes, here we adopt the following parametrized form\cite{2007PhRvD..75d3006H},
 
\beqa
\begin{split}
\frac{d\phi_{\rm \nu}}{dE_{\rm \nu}d\Omega}=&N_0E_{\rm \nu}^{\rm -\gamma-1} \\&\times\left(\frac{a}{1+bE_{\rm \nu}\cos\theta}+\frac{c}{1+eE_{\rm \nu}\cos\theta}\right),
\label{eq:m_ucmh}
\end{split}
\eeqa 
where $\gamma=1.74$, $a = 0.018$, $b = 0.024$, $c = 0.0069$, $e = 0.00139$, $N_0 = 1.95 (1.35) \times 10^{17}$ for neutrinos or antineutrinos. In this work, we set the angle $\theta_{max}=5^\circ $, corresponding to the angular resolution of the neutrino detector, e.g., IceCube, over the relevant energy range, as well as the angle between muons and neutrinos during neutron nucleon scattering~\cite{2007JPhCS..60..334D}.

\begin{figure}[htb]
\centering
\includegraphics[width=0.5\textwidth]{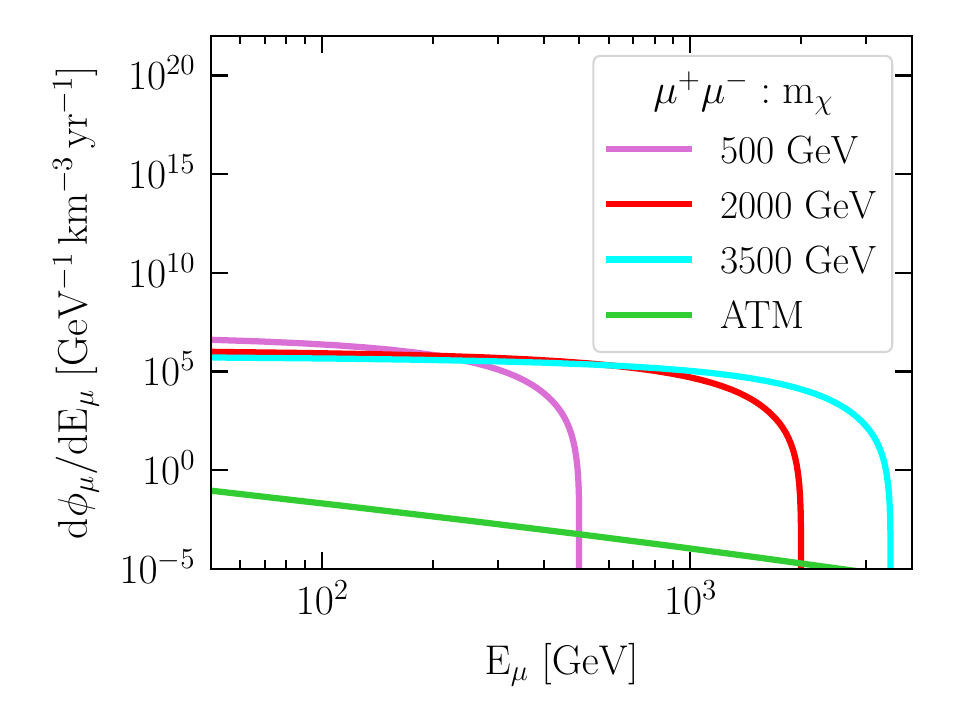}
\hfill
\includegraphics[width=0.5\textwidth]{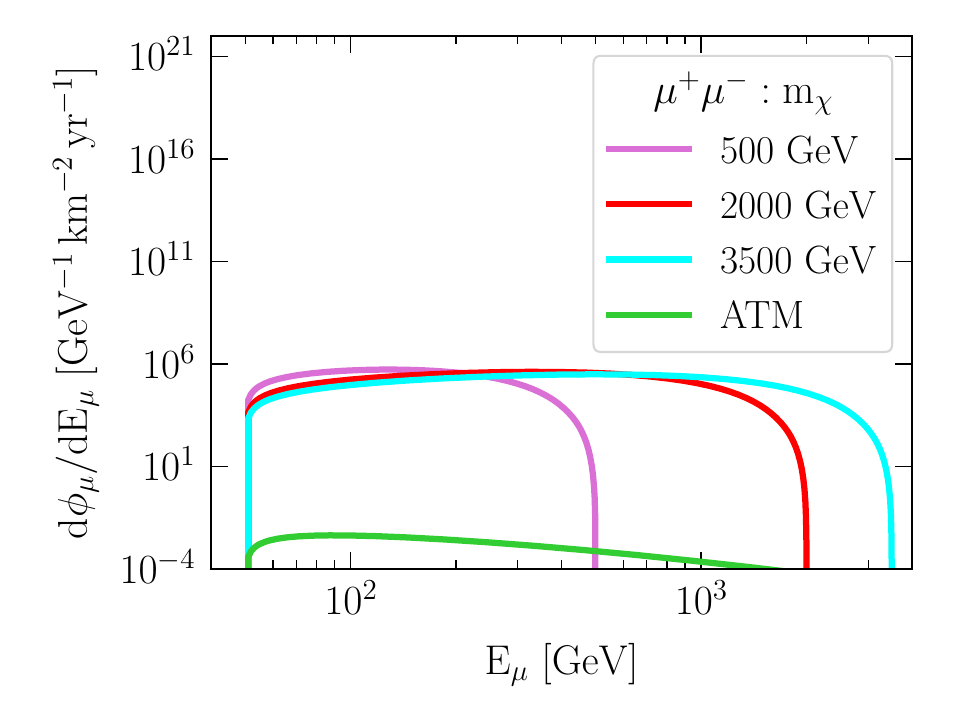}
\caption{The contained events (upper panel) and upward events (down panel) of muon flux from UCMHs due to DM annihilation for $\Omega_{\rm PBH} = 0.1$, DM annihilation channel $\mu^+\mu^-$ and DM mass $m_{\chi} = 500, 2000$ and 3500 GeV. We have set the thermally averaged annihilation cross section of DM $\left<\sigma v\right> =3\times10^{-26}\, \rm cm^3s^{-1}$ as a benchmark value. The atmospheric neutrinos (ATM) are also shown for comparison.}
\label{fig:muon flux}
\end{figure}

Figure~\ref{fig:muon flux} shows the flux of muons for contained (upper panel) and upward (lower panel) events for different DM masses and annihilation channel $\mu^{+}\mu^{-}$ with $\Omega_{\rm PBH}=0.1$, and the atmospheric neutrinos are also shown as the main background for comparison. Depending on the fraction of PBHs, the muon flux from UCMHs due to DM annihilation can exceed the ATM, especially for higher energy (larger DM mass) where the ATM decreases significantly compared with that of lower energy.


\section{constraints on the cosmological abundance of PBHs}
\label{The constraints on the abundance of PBHs}

The number of muon neutrinos ($N_{\rm \nu_{\mu}}$) from extragalactic UCMHs (clothed PBHs) due to DM annihilation can be written as 

\beqa
\begin{split}
N_{\rm \nu_{\mu}, PBHs}=\int_{E_{\rm \mu}^{\rm th}}^{E_{\rm max}}\frac{d\phi_{\rm \mu}}{dE_{\rm \mu}}F_{\rm eff}(E_{\rm \mu})dE_{\rm \mu},
\label{eq:N_PBHs}
\end{split}
\eeqa
where $\frac{d\phi_{\rm \mu}}{dE_{\rm \mu}}$ can be obtained by Eqs.~(\ref{eq:flux_ct}) and (\ref{eq:flux_up}). $F_{\rm eff}(E_{\rm \mu})$ corresponds to the effective volume $V_{\rm eff}$ (effective area $A_{\rm eff}$) of the detector for contained (upward) events. In general, $V_{\rm eff}$ and $A_{\rm eff}$ depend on the energy of detecting particles. Here, for simplicity, we accept that the energy independent effect volume $V_{\rm eff}=0.04\rm km^2$ and the angle-averaged muon effective area $A_{\rm eff}=1 {\rm km}^2 $ for the IceCube experiment~\cite{2009arXiv0907.2263W, 2010PhRvD..81d3508M, 2009NIMPA.602....7R}. In order to obtain the constrains on the cosmological abundance of PBHs, we treat the ATM as the main background, and consider~$\sim 10$ year exposure times for detection. Considering the ATM background, one can obtain the upper limits on the cosmological abundance of PBHs in, e.g., $2\sigma$ statistical significance, using~\cite{Fornengo:2011em,Bergstrom:1997tp} 

\beqa
\begin{split}
\zeta \equiv \frac{N_{\rm PBHs}}{\sqrt{N_{\rm PBHs}+N_{\rm ATM}}},
\label{eq:m_ucmh}
\end{split}
\eeqa
where $N_{\rm ATM}$ is the number of muon neutrinos from ATM and can be obtained via Eq.~(\ref{eq:N_PBHs}).

\begin{figure}
\centering
\includegraphics[width=0.5\textwidth]{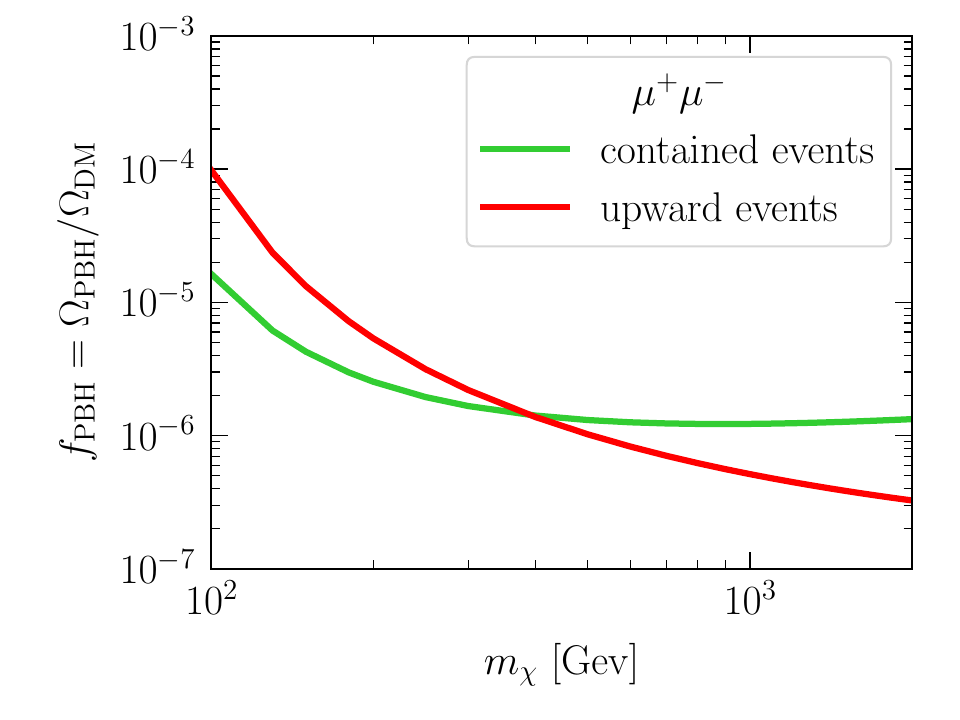}
\caption{The upper limits on the abundance of PBHs $f_{\rm PBH}=\Omega_{\rm PBH}/\Omega_{\rm DM}$ for contained and upward events for the DM annihilation channel $\mu^+\mu^-$, and 10 years exposure time for IceCube experiment. We have set the thermally averaged annihilation cross section of DM $\left<\sigma v\right> =3\times10^{-26}\, \rm cm^3s^{-1}$ as a benchmark value.}
\label{fig:f_PBH}
\end{figure}

The upper limits on the fraction of DM in PBHs, $f_{\rm PBH}=\Omega_{\rm PBH}/\Omega_{\rm DM}$, are shown in Fig.~\ref{fig:f_PBH}. For the PBH mass range considered by us (Eq.~\ref{eq:M_{PBH}}), since we have investigated the extragalactic neutrino flux from clothed PBHs due to DM annihilation, the constraints on $f_{\rm PBH}$ do not depend on the PBHs mass but on the particle DM mass~\cite{PhysRevD.85.125027,2021MNRAS.506.3648C}. For the DM mass $m_{\chi}=100~(1000)~\rm GeV$, the upper limits on the abundance of PBHs are $f_{\rm PBH}=1.6\times 10^{-5}~(1.2\times 10^{-6})$ for contained events and $f_{\rm PBH}=9.8\times 10^{-5}~(5.1\times 10^{-7})$ for upward events, respectively. As shown in Eq.~(\ref{eq:neutrinos_flux}), the differential neutrino flux is larger for smaller DM mass ($d\phi_{\nu}/dE_{\nu}\propto m_{\chi}^{-2}$). However, on the other hand, since we have treated ATM as the main background and it decreases significantly with the increase of energy, as shown in Fig.~\ref{fig:f_PBH}, for the DM mass range considered by us, the final constraints on the abundance of PBHs are stronger for larger DM mass for both contained and upward events. For upward events, the distance $R(E_{\mu})$ is larger for higher energy (corresponding to massive DM particle, Eq.~(\ref{eq:rmu})), resulting in larger muon flux and more events. Therefore, compared with contained events, the constraints on the abundance of PBHs are stronger for larger DM mass ($m_{\chi}\gtrsim 400~\rm GeV$) for upward events. 

\begin{figure}
\centering
\includegraphics[width=0.5\textwidth]{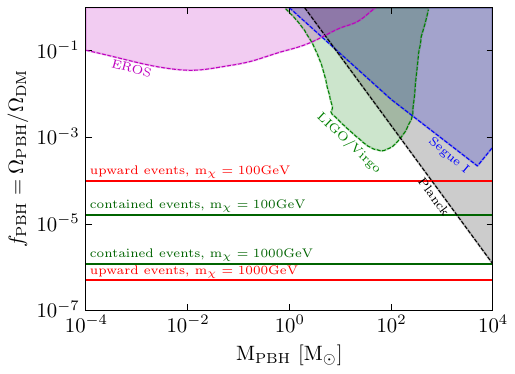}
\caption{The upper limits on the fraction of DM in PBHs, $f_{\rm PBH}=\Omega_{\rm PBH}/\Omega_{\rm DM}$, for several other observations: (i) the merger rate of PBHs in view of the sensitivity of LIGO/Virgo~\cite{PhysRevD.101.043015}; (ii) the dynamical evolution of stars in the dwarf galaxy Segue I due to PBHs~\cite{PhysRevLett.119.041102}; (iii) the influence of accreting PBHs on the CMB (Planck data)~\cite{PhysRevD.96.083524}; (iv) the gravitational lensing effects based on EROS~\cite{EROS-2:2006ryy}. The horizontal solid lines stand for the upper limits obtained in this work for the DM mass $m_{\chi}=100$ and 1000 GeV for contained (dark green) and upward (red) events}
\label{fig:f_PBH_compare}
\end{figure}

The cosmological abundance of PBHs can also be constrained by other different ways (see, e.g., Refs.~\cite{2021RPPh...84k6902C,2022arXiv220514722O,Auffinger:2022khh} for a review). In Fig.~\ref{fig:f_PBH_compare}, we plot the upper limits on the abundance of PBHs as a few other ways for comparison. From this plot, it can be seen that the extragalactic neutrino can provide a useful complement of constraints on the abundance of PBHs for the mass range $M_{\rm PBH}\lesssim 1~M_{\odot}$, where the limits are mainly derived from the gravitational lensing effects based on the observations of the European Southern Observatory (EROS)~\cite{EROS-2:2006ryy}. Note that the constraint on $f_{\rm PBH}$ from the extragalactic gamma-ray background is the strongest one and can be achieved at $f_{\rm PBH}\sim 10^{-10}~(10^{-9})$ for $m_{\chi}=100~(1000)$ GeV~\cite{2021MNRAS.506.3648C}, which is not shown in Fig.~\ref{fig:f_PBH_compare}. 


\section{conclusions}
\label{conclusions}

In the mixed DM scenarios consisting of PBHs and particle DM (WIMPs), PBHs can accrete surrounding WIMPs to form UCMHs after their formation. The number density of DM particles in UMCHs is larger than that of classical DM halos, resulting in a large DM annihilation rate. Moreover, since UCMHs can be formed in the early universe ($z\sim z_{\rm eq}$), it is expected that the WIMPs annihilation in UCMHs can have significant contributions to, e.g., extragalactic gamma-ray/neutrino flux, depending on the fraction of PBHs and the properties of particle DM. The extragalactic gamma-ray flux from clothed PBHs due to DM annihilation has been investigated in previous works and in this work we have studied the neutrino flux. 

There are three flavors of neutrinos, and here we have focused on the muon neutrino. We have investigated the contained and upward events for the purpose of neutrino detection. The differential neutrino flux can exceed the ATM for large abundance of PBHs (e.g., Fig.~\ref{fig:muon flux} for $\Omega_{\rm PBH}=0.1$), especially for higher energy where the ATM decreases significantly. Compared with the ATM, the main neutrino background considered by us, we obtained the upper limits on the fraction of DM in PBHs for IceCube experiment for 10 years of exposure time. For the DM mass $m_{\chi}=100~(1000)$ GeV, the upper limits (2$\sigma$) on the fraction of PBHs are $f_{\rm PBH}=1.6\times 10^{-5}~(1.2\times 10^{-6})$ for contained events and $f_{\rm PBH}=9.8\times 10^{-5}~(5.1\times 10^{-7})$ for upward events, respectively. Many other ways can also be used to constrain the cosmological abundance of PBHs. Compared with other constraints, especially for the extragalactic gamma-ray background from clothed PBHs due to DM annihilation, although the upper limits obtained by us are not the strongest, it is a different way to study the cosmological abundance of PBHs.

\section{Acknowledgements}
This work is supported by the Shandong Provincial Natural Science Foundation (Grant No.ZR2021MA021) and the National Natural Science Foundation of China (Grant No. U2038106). 
\

\newpage 

\bibliographystyle{apsrev4-1}
\bibliography{ref}

\end{document}